\documentclass[10pt,letterpaper]{article}
\usepackage[utf8]{inputenc}
\usepackage[T1]{fontenc}
\usepackage{amsmath}
\usepackage{amsfonts}
\usepackage{amssymb}
\usepackage{amsthm}
\usepackage{wasysym}
\usepackage{graphicx}
\usepackage{csquotes}
\usepackage{microtype}
\usepackage{enumerate}
\usepackage{subcaption}
\usepackage{calc}
\usepackage{booktabs}
\usepackage[svgnames]{xcolor}
\usepackage[labelfont={bf}]{caption}
\usepackage[hidelinks]{hyperref}
\usepackage{cleveref}
\usepackage[citestyle=authoryear,style=authoryear,natbib=true,sorting=nyt]{biblatex}

\renewcommand{\epsilon}{\varepsilon}
\renewcommand{\phi}{\varphi}

\newtheorem{lemma}{Lemma}

\theoremstyle{definition}
\newtheorem{defn}{Definition}

\newtheorem{example}{Example}

\renewcommand{\vec}[1]{\mathbf{#1}}
\newcommand{\mat}[1]{\mathbf{#1}}
\newcommand{\marginnote}[1]{\leavevmode\marginpar{\footnotesize\raggedright\slshape{#1}}}

\definecolor{tabBlue}{HTML}{4E79A7}
\definecolor{tabOrange}{HTML}{F28E2B}
\definecolor{tabRed}{HTML}{E15759}
\definecolor{tabLightBlue}{HTML}{76B7B2}
\definecolor{tabGreen}{HTML}{59A14F}
\definecolor{tabYellow}{HTML}{EDC948}
\definecolor{tabPurple}{HTML}{B07AA1}
\definecolor{tabPink}{HTML}{FF9DA7}
\definecolor{tabBrown}{HTML}{9C755F}
\definecolor{tabGray}{HTML}{BAB0AC}

\author{Andreas Stöckel\\Centre for Theoretical Neuroscience\\University of Waterloo}
\title{Constructing Dampened LTI Systems Generating Polynomial Bases}
\date{March 9, 2021}

\newcommand{\mA}{\mat{\bar A}}
\newcommand{\mB}{\mat{\bar B}}
\newcommand{\nA}{\mat{\hat A}}
\newcommand{\nB}{\mat{\hat B}}

\newcommand{\cPos}[1]{\textbf{\textcolor{DarkBlue}{#1}}}
\newcommand{\cNeg}[1]{\textbf{\textcolor{DarkRed}{#1}}}

\begin{document}
\maketitle

\begin{abstract}
We present an alternative derivation of the LTI system underlying the Legendre Delay Network (LDN).
To this end, we first construct an LTI system that generates the Legendre polynomials.
We then dampen the system by approximating a windowed impulse response, using what we call a \enquote{delay re-encoder}.
The resulting LTI system is equivalent to the LDN system.
This technique can be applied to arbitrary polynomial bases, although there typically is no closed-form equation that describes the state-transition matrix.
\end{abstract}

\section{Introduction}

The Delay Network, originally proposed by \citet{voelker2018improving}, is a recurrent neural network capable of delaying an input signal $u(t)$ by $\theta$ seconds.
\Citet{voelker2019} points out that the impulse response of the linear time-invariant (LTI) system underlying the delay network traces out the shifted Legendre polynomials.
We hence refer to this network as the \emph{Legendre Delay Network} (LDN), and to the LTI system underlying the LDN as the \emph{LDN system}.

The LDN has been derived from the Padé approximants of a Laplace domain delay $e^{-\theta s}$ and a subsequent conditioning coordinate transformation.
From this perspective, the relationship to the Legendre polynomials is rather surprising.

\Citet*{gu2020hippo} have proposed LTI systems similar to the LDN system for other polynomial bases and various window functions.
These systems are derived in the opposite direction of Voelker's original approach.
Given a polynomial basis and a window function, \citeauthor{gu2020hippo} derive an LTI system that realizes this basis with the desired weighting applied.

In this report, we use a similar approach.
Although the method presented here has been developed independently, our approach differs from \citeauthor{gu2020hippo} mostly in terms of presentation.
Our goal is to provide a simple derivation assuming a minimal mathematical background; readers are encouraged to consult \citeauthor{gu2020hippo} for a more general treatment of the topic.

\section{Deriving the LDN system}

\begin{figure}[b]
\setlength\fboxsep{0.5em}
\noindent\colorbox{GhostWhite}{\parbox{\textwidth - 1em}{\small
\textbf{Notation:} Due to an unfortunate shortage of letters in the Latin alphabet, we resort to a somewhat confusing, but hopefully consistent notation.
In particular, we use the following symbols to denote linear time-invariant (LTI) system matrices:\\[0.25cm]
\hspace*{0.2cm}\begin{tabular}{r l}
	\toprule
	\emph{Symbols} & \emph{Description}\\
	\midrule
	$\mat A$, $\mat B$ & General LTI system with feedback matrix $\mat A$ and input matrix $\mat B$\\
	$\nA$, $\nB$ & The scaled Legendre Delay Network (LDN) system\\
	$\nA'$, $\nB'$ & The original LDN system proposed by \citet{voelker2019}\\
	$\mA$, $\mB$, $\mat{\bar \Gamma}$ & The Legendre system and the corresponding delay re-encoder\\
	\bottomrule
\end{tabular}\\[0.25cm]
\noindent\textbf{Source code:}
We provide Python code for many of the equations in this report (see margin notes). This code can be found here:\\[0.125cm]\hspace*{0.5cm}\url{https://github.com/astoeckel/dlop_ldn_function_bases}
}}
\end{figure}

We derive the LDN system $\nA$, $\nB$ in two steps.
First, we construct an LTI system $\mA$, $\mB$ that traces out the Legendre polynomials as its impulse response over the interval $[0, \theta]$.
Second, we derive a matrix $\mat{\bar \Gamma}$ that decodes a delayed signal $u(t - \theta)$ from the state vector $\vec m(t)$ and re-encodes this delayed function in terms of the Legendre basis.
We call this matrix $\mat{\bar\Gamma}$ \enquote{delay re-encoder}.
Subtracting $\mat{\bar\Gamma}$ from $\mA$ results in a dampened impulse response; the LDN system is simply given as $\nA = \mA - \mat{\bar \Gamma}$ and $\nB = \mB$.

\subsection{The LDN System}

Before we discuss an alternative derivation of the LDN system $\nA$, $\nB$, we should first define this system itself.
Note that we discuss a scaled version of the original LDN system. The impulse response of the scaled system matches the shifted Legendre polynomials for $q \to \infty$ (see \Cref{sec:q_to_infty}).

More precisely, compared to the original system, we divide each state dimension $i \in \{1, \ldots, q\}$ by $2i + 1$. This can be easily accomplished by constructing a diagonal matrix $\mat M$ of scaling factors and using this $\mat M$ as a coordinate transformation.
In other words, it holds $\nA = \mat M \nA' \mat M^{-1}$ and $\nB = \mat M \mat \nB'$, where $\nA'$, $\nB'$ is the original LDN system \citep[cf.][Section 6.3.1, pp.~133-135]{voelker2019}.

\begin{defn}[LDN System]
Let $\vec m \in \mathbb{R}^q$, $\nA \in \mathbb{R}^{q \times q}$ and $\nB \in \mathbb{R}^{q \times 1}$.
The scaled LDN system is given as
\begin{align}
	\begin{aligned}
	\frac{\mathrm{d}}{\mathrm{d}t} \theta \vec m(t) &= \nA \vec m(t) + \nB u(t)\\
	\big( \nA \big)_{ij} &= (2j - 1) \begin{cases}
		-1 & \text{if } i \leq j \text{ or } i + j \text{ is even} \,, \\
		 1 & \text{if } i > j \text{ and } i + j \text{ is odd} \,,\\
	\end{cases} \\
	\big( \nB \big)_i &= (-1)^{i + 1} \,.
	\end{aligned}
	\label{eqn:ldn_system}
\end{align}
\marginnote{~~\\[-8.75em]This equation is implemented in the function \texttt{mk\_ldn\_lti}\,.}%
The window-length $\theta$ determines how fast the system evolves.
\end{defn}

\begin{example}
The LDN system for $\theta = 1$ and $q = 6$ is given as
\begin{align*}
	\nA &= \begin{pmatrix}
		\cNeg{-1}  &	\cNeg{-3}  &	\cNeg{-5}  &	\cNeg{-7}  &	\cNeg{-9}  &	\cNeg{-11} \\
		\cPos{1}   &	\cNeg{-3}  &	\cNeg{-5}  &	\cNeg{-7}  &	\cNeg{-9}  &	\cNeg{-11} \\
		\cNeg{-1}  &	\cPos{3}   &	\cNeg{-5}  &	\cNeg{-7}  &	\cNeg{-9}  &	\cNeg{-11} \\
		\cPos{1}   &	\cNeg{-3}  &	\cPos{5}   &	\cNeg{-7}  &	\cNeg{-9}  &	\cNeg{-11} \\
		\cNeg{-1}  &	\cPos{3}   &	\cNeg{-5}  &	\cPos{7}   &	\cNeg{-9}  &	\cNeg{-11} \\
		\cPos{1}   &	\cNeg{-3}  &	\cPos{5}   &	\cNeg{-7}  &	\cPos{9}   &	\cNeg{-11} 
	\end{pmatrix} \,, &
		\nB &= \begin{pmatrix}
			\cPos{1}   \\
			\cNeg{-1}  \\
			\cPos{1}   \\
			\cNeg{-1}  \\
			\cPos{1}   \\
			\cNeg{-1}  
		\end{pmatrix} \,.
\end{align*}
The impulse response of this system is depicted in \Cref{fig:ldn_impulse}.
\end{example}

For finite $q$, the impulse response \emph{approximates} the Legendre polynomials.
In contrast to the Legendre polynomials, the impulse response quickly converges to zero.
In other words, the impulse response can be thought of as a \enquote{dampened} version of the Legendre polynomials.
This almost finite impulse response\footnote{We refer to the impulse response as \enquote{almost finite}, because it decays exponentially for $t > \theta$. Hence, technically, the system does \emph{not} have a finite impulse response. However, this is negligible for all practical purposes.} is the most practically useful property of the LDN system.
Feeding an input signal $u(t)$ into the LDN system can be thought of as performing an online basis transformation.
That is, the state $\vec m(t)$ approximately represents a window of the input signal $u_{[t - \theta, t]}$ in terms of a linear combination of the Legendre polynomials; we discuss this in more detail below.
In general, such transformations are known as \enquote{sliding transformations}.

\subsection{The Legendre System}

Instead of \emph{approximating} the Legendre polynomials, we can construct an LTI system $\mA$, $\mB$ that \emph{perfectly} traces out the Legendre polynomials as its impulse response.
In other words, the system \enquote{generates} the Legendre polynomials.
As we will see in the next subsection, we can then subtract a dampening term $\mat \Gamma$ from this \enquote{Legendre system} to obtain the LDN.

\begin{lemma}
The impulse response of the linear time-invariant system $\frac{d}{dt} \theta \vec m(t) = \mA \vec m(t) + \mB u(t)$ with $\vec m \in \mathbb{R}^q$, $\mA \in \mathbb{R}^{q \times q}$ and $\mB \in \mathbb{R}^{q \times 1}$
\begin{align}
	\begin{aligned}
	\big(\mA\big)_{ij} &= (4j - 2) \begin{cases}
			0 & \text{if } i \leq j \text{ or } i + j \text{ is even} \,, \\
			4j - 2 & \text{if } i > j \text{ and } i + j \text{ is odd} \,,
		\end{cases} &
	\big(\mB\big)_i &= (-1)^{i + 1} \,,
	\end{aligned}
	\label{eqn:legendre_system}
\end{align}
\marginnote{~~\\[-6.25em]This equation is implemented in the function \texttt{mk\_leg\_lti}\,.}%
are the first $q$ shifted Legendre polynomials $\tilde P_n(t \theta^{-1})$ over $t \in [0, \theta]$.
\label{lem:leg_sys}
\end{lemma}

\begin{figure}
	\centering
	\includegraphics{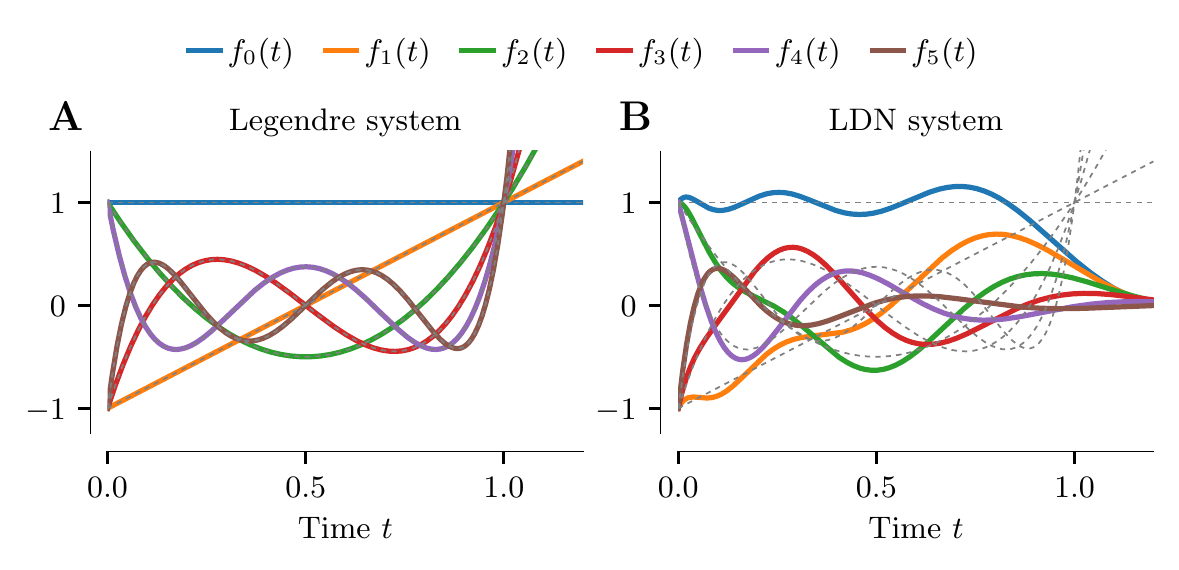}%
	\begin{subfigure}{0.0\textwidth}\phantomcaption\label{fig:leg_impulse}\end{subfigure}%
	\begin{subfigure}{0.0\textwidth}\phantomcaption\label{fig:ldn_impulse}\end{subfigure}%
	\caption{Impulse responses $f_n(t)$ of the Legendre and LDN LTI systems for $\theta = 1$ of order $q = 6$.
	Dashed grey lines correspond to the first six shifted Legendre polynomials $\tilde P_n(t)$.
	\textbf{(A)} Impulse response of the Legendre system.
	The impulse response perfectly traces out the Legendre polynomials. The system diverges for $t \to \infty$.
	\textbf{(B)} Impulse response of the LDN system.
	The system quickly converges to zero for $t \to \infty$.}
	\label{fig:ldn_leg_impulse}
\end{figure}

\begin{example}
The LTI system constructing the first six shifted Legendre polynomials as its impulse response for $\theta = 1$ is given as
\begin{align*}
	\mA &= \begin{pmatrix}
		      0    &	      0    &	      0    &	      0    &	      0    &	      0    \\
		\cPos{2}   &	      0    &	      0    &	      0    &	      0    &	      0    \\
		      0    &	\cPos{6}   &	      0    &	      0    &	      0    &	      0    \\
		\cPos{2}   &	      0    &	\cPos{10}  &	      0    &	      0    &	      0    \\
		      0    &	\cPos{6}   &	      0    &	\cPos{14}  &	      0    &	      0    \\
		\cPos{2}   &	      0    &	\cPos{10}  &	      0    &	\cPos{18}  &	      0    
	\end{pmatrix} \,, &
	\mB &= \begin{pmatrix}
		\cPos{1}   \\
		\cNeg{-1}  \\
		\cPos{1}   \\
		\cNeg{-1}  \\
		\cPos{1}   \\
		\cNeg{-1}  
	\end{pmatrix} \,.
\end{align*}
The impulse response of this system is depicted in \Cref{fig:leg_impulse}.
\end{example}

\begin{proof}[Proof of \Cref{lem:leg_sys}]
As pointed out by \citet[Appendix B.1.1]{gu2020hippo}, and originally described in \citet[Chapter 12.2, p.~751]{arfken2005mathematical}, Legendre polynomials fulfil the following recurrence relation with respect to their derivative:
\begin{align*}
	\frac{\mathrm{d}}{\mathrm{d}t} \big( P_{n + 1}(t) - P_{n - 1}(t) \big) &= (2n + 1) P_n(t) \,.
\end{align*}
Substituting in $\tilde P_n(t \theta^{-1}) = P_n((2t - 1) \theta^{-1})$ and rearranging we get
\begin{align*}
	\frac{\mathrm{d}}{\mathrm{d}t} \theta \tilde P_{n + 1}(t \theta^{-1})
		&= (4n + 2) \tilde P_n(t \theta^{-1}) + \frac{\mathrm{d}}{\mathrm{d}x} \tilde P_{n - 1}(t \theta^{-1}) \\
		&= (4n + 2) \tilde P_n(t \theta^{-1}) + (4(n - 2) + 2) \tilde P_{n - 2}(t \theta^{-1}) + \ldots \,.
\end{align*}
This recurrence relation terminates with $\tilde P_0$ or $\tilde P_1$ depending on whether $n$ is even or odd.
Crucially, this recurrence relation implies that the differential of the $n$th Legendre polynomial can be expressed as a linear combination of the preceding Legendre polynomials.
Let $\vec m(t) = (\tilde P_0(t \theta^{-1}), \ldots, \tilde P_{q - 1}(t \theta^{-1}))$. We can now write the above equation as a vector-matrix product
\begin{align*}
	\frac{\mathrm{d}}{\mathrm{d}t} \theta \vec m(t) &= \mA \vec m(t) \,,
\end{align*}
where $\mA$ is as defined in \cref{eqn:legendre_system}.
The vector $\mB = \big( \tilde P_0(0), \ldots, \tilde P_{q - 1}(0)\big)$ defines the initial value of each state dimension in response to a Dirac pulse $u(t) = \delta(t)$.
\end{proof}

\paragraph{Sliding transformations}
Feeding a signal $u(t)$ into the Legendre system convolves this signal with the Legendre polynomials.
The state vector $\vec m(t) = (m_0(t), \ldots, m_{q - 1}(t))$ is the convolution between the impulse response $\exp(\mA t) \mB = \tilde P_n(t \theta^{-1})$ and the input signal $u(t)$:
\begin{align}
	m_n(t)
		&= \int_{0}^t \big( \exp(\mA \tau) \mB \big)_{n + 1} u(t - \tau) \,\mathrm{d}\tau
		 = \int_{0}^t \tilde P_n(\tau \theta^{-1}) u(t - \tau) \,\mathrm{d}\tau \,.
	\label{eqn:ldn_basis_trafo}
\end{align}
At $t = \theta$ a segment of the input $u_{[0, \theta]}$ is stored in $\vec m(\theta)$ as a linear combination of the basis functions $\tilde P_n$.
At this point, the convolution operator is mathematically equivalent to the inner product between the Legendre polynomials and the input $u(t)$ up to $t = \theta$, i.e., $m_n(\theta) = \langle u_{[0, \theta]}, \tilde P_n \rangle$.
However, for $t > \theta$, the Legendre polynomials $P_n(t \theta^{-1})$ and our impulse response quickly diverge.

If we were able to limit the impulse response to a window $[0, \theta]$ (i.e., $t > \theta \Rightarrow \exp(\mat A \tau)\mat B = 0$), then the state vector $\vec m(t)$ would always represent a slice of the most recent input history window $u_{[t - \theta, t]}$.
This suggests a convenient technique for implementing a \enquote{sliding transformation}.
Advancing the LTI system computes the spectral coefficients $\vec m(t)$ describing a recent input slice.
Using more general notation, we would optimally like to implement a windowed convolution
\begin{align}
	m_n(t)
		&= \int_{0}^\theta f_n(\tau) u(t - \tau) \,\mathrm{d}\tau
		 = \langle f_n, u_{[t - \theta, t]} \rangle \,,
	\label{eqn:ldn_basis_trafo_window}
\end{align}
where the $f_n$ are the desired basis functions over $[0, \theta]$ generated by the LTI system $\mat A$, $\mat B$, i.e., $f_n(t) = (\exp(\mat A t) \mat B)_{n +1}$ for $n \in \{0, \ldots, q - 1\}$.

\subsection{Dampening Through Information Erasure}

To enforce a finite impulse response---to \emph{dampen} the system---we somehow need to prevent the impulse response from evolving past $\theta$.
We first discuss how to accomplish this under the assumption that we have access to a perfect delay---i.e., we have a recording of the input signal over the past $\theta$ seconds.
In a second step we approximate a perfect delay by decoding a delayed version of the input signal from the system state $\vec m(t)$ itself.

\paragraph{Constructing a rectangle window using a perfect delay}
If we have access to a perfect delay line of length $\theta$, we can easily construct a system with a finite impulse response.
All we need to do this, is to subtract the delayed $u(t - \theta)$ from the system state $\vec m(t)$ using an \enquote{encoding vector} $\vec e(\theta)$.
Put differently, we erase information about $u(t)$ older than $\theta$ seconds from the system state.
We hence refer to this method as \enquote{information erasure}.

\begin{lemma}
\label{lem:rectangle_window}
Let $\mat A$, $\mat B$ describe an LTI system and let $u(t)$ be some input signal.
The impulse response of the following modified system
is unchanged compared to the original LTI system for $0 \leq t < \theta$ but zero for all $t \geq \theta$%
\begin{align}
	\frac{\mathrm{d}}{\mathrm{d}t} \vec m(t) &= \mat A \vec m(t) + \mat B u(t) - \vec e(\theta) u(t - \theta) \,, & \text{where } \vec e(\theta) = \exp(\mat A \theta) \mat B \,.
	\label{eqn:information_erasure}
\end{align}
\begin{proof}[Proof of \Cref{lem:rectangle_window}]
Consider the impulse response, i.e., $u(t) = \delta(t)$, where $\delta(t)$ is a Dirac pulse.
The system state is unchanged for $t < \theta$ as $\delta(t - \theta) = 0$ for all $t \neq \theta$; hence the impulse response of the system is $\vec m(t) = \exp(\mat A t) \mat B$ for $0 \leq t < \theta$.
At $t = \theta$, according to the definition of the Dirac pulse, we subtract $\vec e(\theta) = \exp(\mat A \theta) \mat B$ from the system state $\vec m(\theta)$, exactly when $\vec m(\theta)$ is equal to $\vec e(\theta)$.
The resulting $\vec m$ is zero and remains zero as $u(t) = \delta(t) = 0$ for  $t \neq 0$.
\end{proof}
\end{lemma}

\paragraph{Approximating a delayed input signal}
In practice, we may not have access to the delayed input signal $u(t - \theta)$.
However, remember that our goal is to construct a system such that the system state $\vec m(t)$ represents a slice of $u$ over the interval $[t - \theta, t]$.
This representation is in terms of a linear combination of the $q$ polynomial basis functions.
We can thus decode an approximate $u(t - \theta')$ from the state $\vec m(t)$ using a delay decoder $\vec d(\theta')$.

\begin{defn}[Delay decoder]
Let $\hat u : [0, \theta] \longrightarrow \mathbb{R}$ be a linear combination of $q$ basis functions $f_n : [0, \theta] \longrightarrow \mathbb{R}$ with weighting coefficients $\xi_n$.
Let furthermore $\vec m(t) = (m_0(t), \ldots, m_{q - 1}(t))$ denote the state of a $q$-dimensional linear dynamical system with impulse responses $f_n$ and input $\hat u$ (cf.~eq.~\ref{eqn:ldn_basis_trafo_window}), i.e.,
\begin{align*}
	m_n(t)
		&= \int_{0}^t f_n(\tau) \hat u(t - \tau) \,\mathrm{d}\tau
		 = \int_{0}^t f_n(\tau) \sum_{m = 0}^{q - 1} \xi_m f_m(t - \tau )\,\mathrm{d}\tau \,.
\end{align*}
Then $\vec d(\theta') = (d_0(\theta'), \ldots, d_{q - 1}(\theta'))$ is called a \emph{delay decoder} if
\begin{align}
	\begin{aligned}
	\hat u(\theta - \theta')
		 &= \sum_{m = 0}^{q - 1} \xi_m f_m(\theta - \theta')
		 = \big\langle \vec d(\theta'), \vec m(\theta) \big\rangle \\
		 &= \sum_{n = 0}^{q - 1} d_n(\theta') \int_{0}^\theta f_n(\tau) \sum_{m = 0}^{q - 1} \xi_m f_m(\theta - \tau )\,\mathrm{d}\tau \,.
	\end{aligned}
	\label{eqn:delay_decoder}
\end{align}
\end{defn}

\begin{lemma}
For the Legendre polynomials $\tilde P_n(t \theta^{-1})$, the delay decoder $\vec d(\theta')$ is
\begin{align*}
	\big( \vec d(\theta') \big)_m &= \frac{2m + 1}{\theta} \tilde P_m(\theta' \theta^{-1}) \,.
\end{align*}
\begin{proof}
	Let $f_n(t) = \tilde P_n(t \theta^{-1})$. Combining the proposed delay decoder $\vec d(\theta')$ with the delay decoder definition yields
	\begin{align*}
		\big\langle \vec d(\theta'), \vec m(\theta) \big\rangle
			&= \sum_{n = 0}^{q - 1} \frac{2n + 1}{\theta} \tilde P_n(\theta' \theta^{-1}) \int_{0}^\theta \tilde P_n(\tau \theta^{-1}) \sum_{m = 0}^{q - 1} \xi_m \tilde P_m((\theta - \tau )\theta^{-1})\,\mathrm{d}\tau \\
			&= \sum_{n = 0}^{q - 1} \sum_{m = 0}^{q - 1} \xi_m \frac{2n + 1}{\theta} \tilde P_n(\theta' \theta^{-1}) \int_{0}^\theta \tilde P_n(\tau \theta^{-1}) \tilde P_m((\theta - \tau )\theta^{-1})\,\mathrm{d}\tau \,.
	\end{align*}
	The integral can be simplified using two properties of the Legendre polynomials
	\begin{align*}
		\int_0^1 \tilde P_n(\tau) \tilde P_m(\tau) \, \mathrm{d}\tau &= \frac{\delta_{nm}}{2m + 1} \,, & \text{(Orthogonality)}  \\
		\tilde P_n(t) &= (-1)^n \tilde P_n(1 - t) \,, & \text{(Parity)}
	\end{align*}
	where $\delta_{ij}$ is the Kronecker delta. Continuing the above set of equations we get
	\begin{align*}
		\big\langle \vec d(\theta'), \vec m(\theta) \big\rangle
			&= \sum_{n = 0}^{q - 1} \sum_{m = 0}^{q - 1} \xi_m \frac{2n + 1}{\theta} \tilde P_n(\theta' \theta^{-1}) (-1)^n \frac{\theta \delta_{mn} }{2n + 1} \hspace{2.97cm}\\
			&= \sum_{n = 0}^{q - 1} \xi_n (-1)^n \tilde P_n(\theta' \theta^{-1})
			 = \hat u(\theta - \theta') \,. \qedhere
	\end{align*}
\end{proof}
\end{lemma}

Given the concept of a \enquote{delay decoder} we can construct an approximate version of eq.~(\ref{eqn:information_erasure}) that reconstructs $u(t - \theta)$ from the system state $\vec m(t)$
\begin{align}
	\frac{\mathrm{d}}{\mathrm{d}t} \vec m(t) &= \mat A \vec m(t) + \mat B u(t) - \vec e(\theta) \langle \vec d(\theta),  \vec m(t) \rangle \,.
	\label{eqn:information_erasure_approx}
\end{align}
The approximate nature of this equation comes from $u(t)$ not necessarily being expressible as a linear combination of the basis functions, as required in our definition of \enquote{delay decoder}.
We discuss the consequences of this in more detail in \Cref{sec:q_to_infty}; before we get there, we first simplify \cref{eqn:information_erasure_approx} a little further by collapsing the encoder $\vec e(\theta)$ and the delay decoder $\vec d(\theta)$ into a single matrix $\mat \Gamma$.

\begin{defn}[Delay re-encoder]
Let $\vec e(\theta)$ be the encoding vector containing the impulse response of a linear system $\mat A$, $\mat B$ at time $\theta$, i.e.,
\begin{align}
	\vec e(\theta) &= \exp(\mat A \theta) \mat B = \big( f_0(\theta), \ldots, f_{q - 1}(\theta) \big)\,,
	\label{eqn:encoder}
\end{align}
and $\vec d(\theta)$ denote the delay decoder of order $q$ for the basis functions $f_n$ produced by the impulse response of the dynamical system.
Furthermore, let \enquote{$\odot$} denote the outer product.
Then $\mat \Gamma = \vec e(\theta) \odot \vec d(\theta)$ with $\mat \Gamma \in \mathbb{R}^{q \times q}$ is the \emph{delay re-encoder} of order $q$.
This matrix can be used to write \cref{eqn:information_erasure_approx} more compactly
\begin{align*}
	\frac{\mathrm{d}}{\mathrm{d}t} \vec m(t) &= \big( \mat A - \mat \Gamma \big) \vec m(t) + \mat B u(t)
\end{align*}
\end{defn}

\begin{example}
For the shifted Legendre polynomials (and, correspondingly, the Legendre system) the delay re-encoder $\mat{\bar \Gamma}$ is simply given as
\begin{align}
	\big( \mat{\bar \Gamma} \big)_{ij} &= e_i(\theta) d_j(\theta) = \frac{2j + 1}{\theta} \tilde P_{i - 1}(\theta \theta^{-1}) \tilde P_{j - 1}(\theta \theta^{-1}) = \frac{2j + 1}{\theta} \,.
	\label{eqn:legendre_delay_reencoder}
\end{align}
For $\theta = 1$ and $q = 6$ the delay re-encoder is
\begin{align*}
	\mat{\bar \Gamma} = 	
	\begin{pmatrix}
		\cPos{1}   &	\cPos{3}   &	\cPos{5}   &	\cPos{7}   &	\cPos{9}   &	\cPos{11}  \\
		\cPos{1}   &	\cPos{3}   &	\cPos{5}   &	\cPos{7}   &	\cPos{9}   &	\cPos{11}  \\
		\cPos{1}   &	\cPos{3}   &	\cPos{5}   &	\cPos{7}   &	\cPos{9}   &	\cPos{11}  \\
		\cPos{1}   &	\cPos{3}   &	\cPos{5}   &	\cPos{7}   &	\cPos{9}   &	\cPos{11}  \\
		\cPos{1}   &	\cPos{3}   &	\cPos{5}   &	\cPos{7}   &	\cPos{9}   &	\cPos{11}  \\
		\cPos{1}   &	\cPos{3}   &	\cPos{5}   &	\cPos{7}   &	\cPos{9}   &	\cPos{11}  
	\end{pmatrix}
	 \,.
\end{align*}
Notice that, in general, $\nA = \mA - \mat{\bar \Gamma}$.
That is, the LDN feedback matrix $\nA$ (eq.~\ref{eqn:ldn_system}) is simply the difference between the Legendre system feedback matrix $\mA$ (eq.~\ref{eqn:legendre_system}) and the delay re-encoder $\mat{\bar \Gamma}$ for the Legendre system (eq.~\ref{eqn:legendre_delay_reencoder}).
\end{example}

\subsection{Impulse response in the limit $q \to \infty$}
\label{sec:q_to_infty}

The \enquote{information erasure} technique from  \cref{eqn:information_erasure_approx} uses a delay decoder $\vec d(\theta')$ to reconstruct a delayed input signal $u(t - \theta')$ from the system state $\vec m(t)$.
In general, this reconstruction is an approximation, since most $u(t)$ cannot be expressed as a linear combination of the basis functions $f_n$ generated by the LTI system---a condition we assumed to be true when we derived $\vec d(\theta')$.

\paragraph{Aliasing of $u(t)$ onto $\hat u(t)$}
Fortunately, this is less of a problem as it may seem---convolving with the impulse response automatically represents the signal $u(t)$ with respect to $f_n$.
To see why this is, consider the case of a \emph{perfectly windowed} LTI system generating a function basis (using \cref{eqn:information_erasure} from \Cref{lem:rectangle_window}).
The system continuously maps (\enquote{aliases}) a segment of the input signal $u_{[t - \theta, t]}$ onto a signal $\hat u_{[t - \theta, t]}$.
This $\hat u_{[t - \theta, t]}$ is a linear combination of the basis functions $f_n$ generated by the LTI system and is exactly the function obtained when applying the delay decoder $\vec d(\theta')$ to the system state $\vec m(t)$
\begin{align*}
	\hat u_{[t - \theta, t]}(\theta')
		&= \langle \vec d(\theta'), \vec m(t) \rangle
		= \sum_{n = 0}^{q - 1} \xi_n f_n(\theta - \theta') \quad \text{for } 0 \leq \theta' \leq \theta\,.
\end{align*}
Note that this aliasing process merely discards components of the input signal that cannot be represented in the function basis---there are no destructive aliasing artefacts as those encountered when violating the Nyquist-Shannon theorem.

\paragraph{Aliasing of an impulse input}
As we saw in \Cref{fig:ldn_leg_impulse}, the impulse response of the LDN system---i.e., the Legendre system with a delay re-encoder (eq.~\ref{eqn:information_erasure_approx})---does not abruptly disappear for $t > \theta$.
Furthermore, the impulse response no longer perfectly traces out the Legendre polynomials.

Correspondingly, functions are no longer represented with respect to the originally desired function basis (e.g., the shifted Legendre polynomials), but with respect to the actual impulse response of the system.
While we derive the delay re-encoder under the assumption that the system generates a desired function basis, subtracting the delay re-encoder from the feedback matrix inevitably causes this assumption to be violated.

Hence, there is no guarantee that this method will actually generate an (almost) finite impulse response.
For example, this technique will not work properly for periodic bases such as the Fourier series, as $\vec d(\theta) = \vec d(0)$ in this case.

The discrepancy in the impulse responses stems from aliasing of the impulse input $u(t) = \delta(t)$.
There is no finite LTI system that can produce a delayed Dirac pulse $\delta(t - \theta)$; in fact, the aliased function $\hat u(t)$ is likely non-zero almost everywhere and has finite energy at $t = 0$.
The former discrepancy results in the ringing artefacts in the LDN system response (\cref{fig:ldn_impulse}), the latter discrepancy in the only gradually disappearing impulse response for $t > \theta$.

\begin{figure}
	\includegraphics{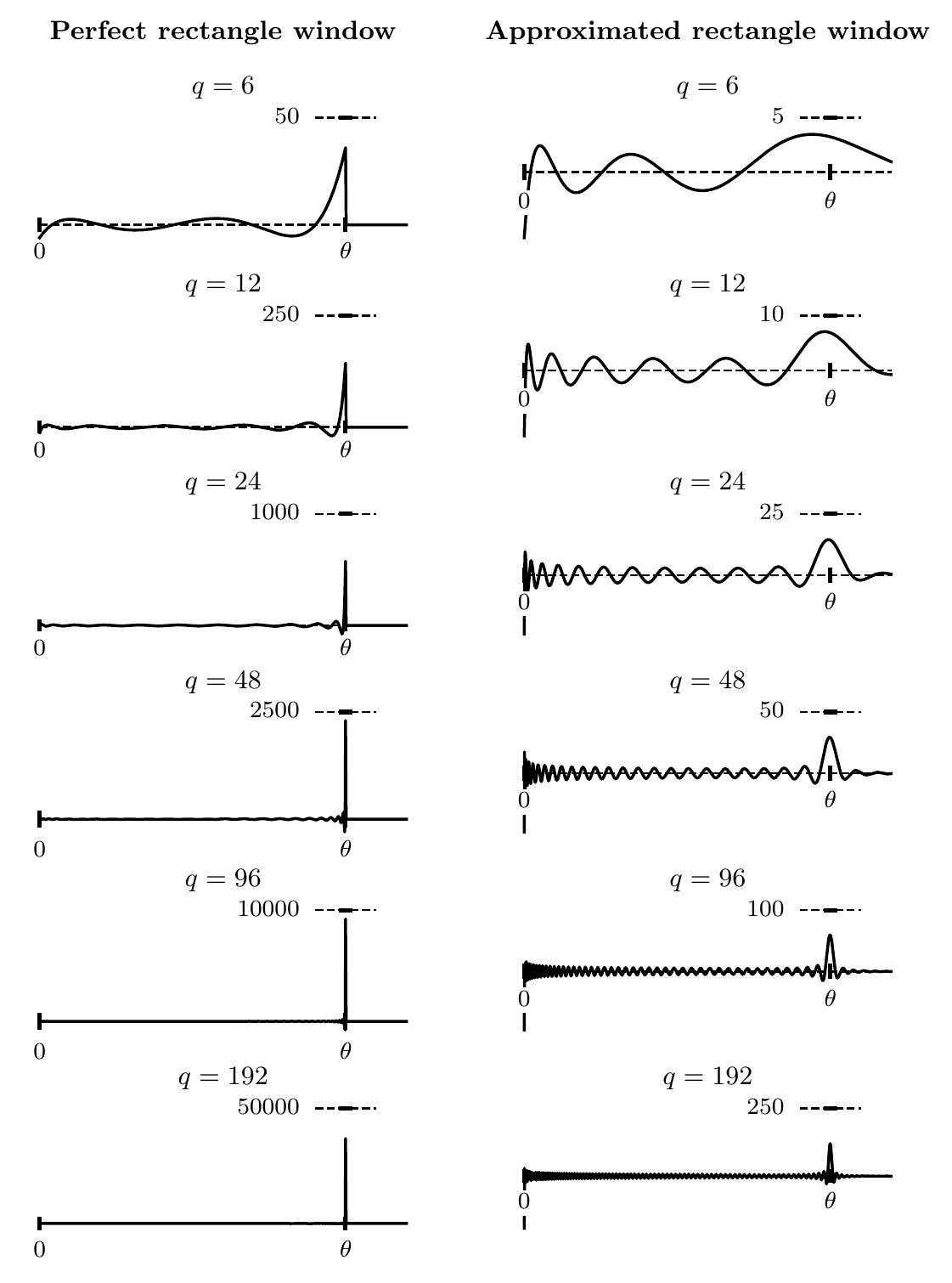}
	\caption{Decoding a delayed impulse input $u(t) = \delta(t)$ for the Legendre/LDN system with different basis function counts $q$.
	Individual graphs depict $\hat u_{[t - \theta, t]}(\theta) = \langle \vec d(\theta), \vec m(t) \rangle$, where $\vec d(\theta)$ is the delay decoder and $\vec m(t)$ is the impulse response of the LTI system.
	\emph{Left:} Delay decoder applied to the impulse response of the Legendre system $\mA$, $\mB$ with a perfect window (eq.~\ref{eqn:information_erasure}). \emph{Right:} Delay decoder applied to the LDN system $\nA$, $\nB$.
	The decoded function gradually converges to a Dirac delta $\delta(t - \theta)$ in both cases, though convergence is slower for the LDN.}
	\label{fig:ldn_leg_impulse_response_decoded}
\end{figure}

The aliased impulses are depicted in \Cref{fig:ldn_leg_impulse_response_decoded} for the Legendre system (with a perfect window applied) and the LDN system (i.e., the Legendre system with delay re-encoder $\mat{\bar \Gamma}$ applied).
Although we do not have a rigorous proof for this, it is safe to assume that in both cases the impulse response converges to a delayed Dirac pulse $\delta(t - \theta)$ for $q \to \infty$; in this case the LDN system and the perfectly windowed Legendre system are exactly equal.

\vspace{0.125cm}

\section{Arbitrary Polynomial Bases}
\label{sec:arbitrary_bases}

The techniques we developed in the previous section can be applied to arbitrary polynomial bases, and not just the Legendre polynomials.
In particular, consider $q$ polynomial basis functions $p_n$ of order $q$
\begin{align*}
	f_n(t) &= p_n(t) = \sum_{k = 0}^{q - 1} \lambda_{n, k} t^k  & \text{for } n \in \{0, \ldots, q - 1\}\,.
\end{align*} 
As we will see, it is quite trivial to construct a $q$-dimensional LTI system that generates the polynomials $p_n$ as its impulse response.
Unfortunately, computing the delay decoder $\vec d(\theta')$ tends to be numerically unstable.

\subsection{Constructing an LTI system generating a basis}

The approach used to prove Lemma 1 can be applied to any set of polynomial basis functions that fulfils two very mild conditions.
First, the polynomials must be linearly independent.
This condition is automatically fulfilled if the functions are indeed chosen from a single function basis.
Second, there must be no $t \in [0, \theta)$ such that $p_n(t) = 0$ for all $n \in \{0, \ldots, q - 1\}$---otherwise the impulse response would be extinguished at that point.

\paragraph{Solving for differentials}
The LTI system feedback matrix $\mat A$ maps the system state $\vec m(t)$ onto the differential $\frac{\mathrm{d}}{\mathrm{d}t} \vec m(t)$.
For an impulse input, the state is supposed to be equal to the polynomials, i.e., $\vec m(t) = (p_0(t), \ldots, p_{q - 1}(t))$.
Furthermore, the differential of a polynomial is just another polynomial. It holds
\begin{align*}
	\frac{\mathrm{d}}{\mathrm{d}t} p_n(t)
		&= \frac{\mathrm{d}}{\mathrm{d}t} \sum_{k = 0}^{q - 1} \lambda_{n, k} t^k
		 = \sum_{k = 1}^{q - 1} k \lambda_{n, k} t^{k-1} \,.
\end{align*}
We hence need to solve for an $\mat A$ that linearly combines the polynomials $p_n$ to form the individual derivatives:
\begin{align}
	\begin{pmatrix}
		\frac{\mathrm{d}}{\mathrm{d}t} p_0(t) \\
		\vdots \\
		\frac{\mathrm{d}}{\mathrm{d}t} p_{q-1}(t)
	\end{pmatrix} &= \mat A
	\begin{pmatrix}
		p_0(t) \\
		\vdots \\
		p_{q-1}(t)
	\end{pmatrix} \,.
	\label{eqn:polynomial_feedback}
\end{align}
Linearly combining two polynomials simply generates a new polynomial where each coefficient $k$ is a weighted sum of the orginal polynomial coefficients.
We can hence write \cref{eqn:polynomial_feedback} as as system of linear equations and solve for $\mat A$
\begin{align*}
	\sum_{m = 0}^{q - 1} a_{m, k} \lambda_{m, k} &= \begin{cases}
		k \lambda_{n, k - 1} & \text{if } k > 0 \,,\\
		0 & \text{if } k = 0 \,,
	\end{cases}
	&
	\begin{aligned}
	 \text{for } & n, k \in \{0, \ldots, q - 1\} \,,\\
	 \text{ and where } &\big(\mat A\big)_{ij} = a_{i - 1, j - 1} \,.
	\end{aligned}
\end{align*}
\marginnote{~~\\[-7em]This equation is implemented in the function \texttt{mk\_poly\_basis\_lti}\,.}%
The input matrix $\mat B$ is the $x$-intercept of the polynomials, i.e., $\big(\mat B\big)_i = \lambda_{i - 1, 0}$.

\begin{figure}
	\centering
	\includegraphics{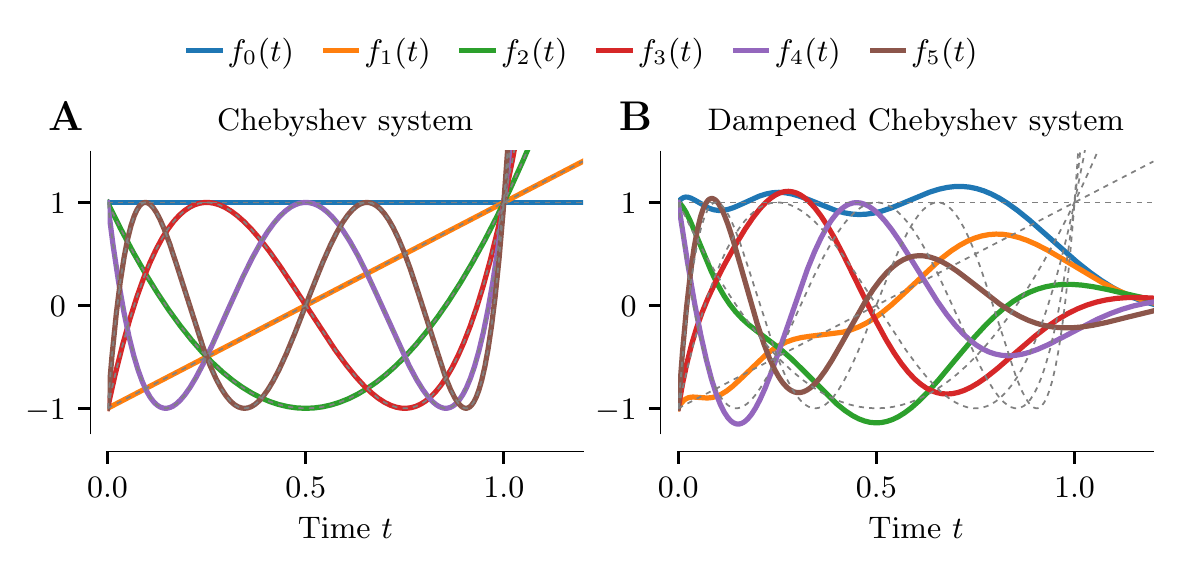}%
	\begin{subfigure}{0.0\textwidth}\phantomcaption\label{fig:cheb_impulse}\end{subfigure}%
	\begin{subfigure}{0.0\textwidth}\phantomcaption\label{fig:cheb_damp_impulse}\end{subfigure}%
	\caption{Impulse responses $f_n(t)$ of the Chebyshev system and the corresponding dampened version ($\theta = 1$ and $q = 6$).
	Dashed grey lines correspond to the first six shifted Chebyshev polynomials.
	\textbf{(A)} Impulse response of the Chebyshev system.
	\textbf{(B)} Impulse response of the dampened Chebyshev system.}
	\label{fig:chebyshev_impulse_response}
\end{figure}

\begin{example}[The Chebyshev system]
The above method can be used to determine the LTI system generating the Chebyshev polynomials.
The following is an extrapolation of numerical results; we did not prove that these equations are indeed correct.
\begin{align*}
	\big(\mat A \big)_{ij} &= (2i - 2) \begin{cases}
			0 & \text{if } i \leq j \text{ or } i + j \text{ is even} \,,\\
			1 & \text{if } j = 1 \text{ and }i > j \text{ and } i + j \text{ is odd} \,,\\
			2 & \text{if } j > 1 \text{ and }i > j \text{ and } i + j \text{ is odd} \,, \\
		\end{cases} &
	\big(\mat B \big)_i &= (-1)^{i + 1} \,.
\end{align*}
\marginnote{~~\\[-8em]These equations are implemented in \texttt{mk\_cheb\_lti}. These matrices can also be approximated using the function \texttt{mk\_cheb\_poly\_basis} and passing the result into \texttt{mk\_poly\_basis\_lti}\,.}%
For $q = 6$ the matrix $\mat A$ takes the following shape
\begin{align*}
\mat{A} &= \begin{pmatrix}
	      0    &	      0    &	      0    &	      0    &	      0    &	      0    \\
	\cPos{2}   &	      0    &	      0    &	      0    &	      0    &	      0    \\
	      0    &	\cPos{8}   &	      0    &	      0    &	      0    &	      0    \\
	\cPos{6}   &	      0    &	\cPos{12}  &	      0    &	      0    &	      0    \\
	      0    &	\cPos{16}  &	      0    &	\cPos{16}  &	      0    &	      0    \\
	\cPos{10}  &	      0    &	\cPos{20}  &	      0    &	\cPos{20}  &	      0    
\end{pmatrix} \,, & \mat{B} &= \begin{pmatrix}
	\cPos{1}   \\
	\cNeg{-1}  \\
	\cPos{1}   \\
	\cNeg{-1}  \\
	\cPos{1}   \\
	\cNeg{-1}  
\end{pmatrix} \,.
\end{align*}
The Chebyshev polynomials and the impulse response of the corresponding LTI system are depicted in \Cref{fig:cheb_impulse}.
\end{example}

\subsection{Computing the delay re-encoder}
To compute the delay re-encoder $\mat \Gamma = \vec e(\theta) \odot \vec d(\theta)$ we need both the encoder $\vec e(\theta)$ and the delay-decoder $\vec d(\theta')$.
While the encoder can be easily computed according to \cref{eqn:encoder}, obtaining $\vec d(\theta)$ is a little trickier.

We present three methods for computing the delay decoder $\vec d(\theta)$.
The first method directly solves for $\vec d(\theta)$; the second method inverts the polynomial basis.
Both methods are unstable for large basis function counts $q$.
The third method is to discretize the polynomials and to use a matrix pseudo-inverse to approximate an inverse of the polynomial basis, which can circumvent some of the instabilities.

\paragraph{Directly solving for $\vec d(\theta)$}
Taking the definition of the delay re-encoder $\vec d(\theta)$ from \cref{eqn:delay_decoder} and substituting $f_n$ with a polynomial basis we obtain
\begin{align*}
	\sum_{i = 0}^{q - 1} d_i(\theta') m_i(\theta)
		&= \sum_{i = 0}^{q - 1} d_i(\theta') \int_0^\theta f_i(\tau) \hat u(\theta - \tau)\,\mathrm{d}\tau \\
		&= \sum_{i = 0}^{q - 1} d_i(\theta') \int_0^\theta f_i(\tau) \sum_{j = 0}^{q - 1} \xi_j f_j(\theta - \tau) \,\mathrm{d}\tau \\
		&= \sum_{i = 0}^{q - 1} d_i(\theta') \int_0^\theta \left( \sum_{k=0}^{q - 1} \lambda_{i, k} \tau^k \right)\sum_{j = 0}^{q - 1} \xi_j \left( \sum_{k=0}^{q - 1} \lambda_{j, k} (\theta - \tau)^k \right) \,\mathrm{d}\tau \\
		&= \sum_{j = 0}^{q - 1} \xi_j \sum_{i = 0}^{q - 1} d_i(\theta') \int_0^\theta \left( \sum_{k=0}^{q - 1} \lambda_{i, k} \tau^k \right) \left( \sum_{k=0}^{q - 1} \lambda_{j, k} (\theta - \tau)^k \right) \,\mathrm{d}\tau \\
		&\overset{!}= \sum_{j = 0}^{q - 1} \xi_j \sum_{k=0}^{q - 1} \lambda_{j, k} (\theta - \theta')^k \quad \text{for all } \xi_j \in \mathbb{R} \\
		&= \hat u(\theta - \theta') \,,
\end{align*}
Substituting $p_j(\theta - \tau)$ with a polynomial $p_j'(\theta) = p_j(\theta - \tau)$ with coefficients $\rho_{j, k}$:
\begin{align*}
	\sum_{k=0}^{q - 1} \lambda_{j, k} (at + b)^k
		&= \sum_{k=0}^{q - 1} \underbrace{\left(\sum_{n=k}^{q - 1} \binom{n}{k} a^k b^{n-k} \lambda_{j, n} \right)}_{\rho_{j, k}}  t^k
		= \sum_{k=0}^{q - 1} \rho_{j, k} t^k \,,
\end{align*}
where $a = -1$, $b = \theta$.
Since the above equality holds for all possible input signals $\hat u$ (i.e., any combination of $\xi_j$) we get for all $j \in \{0, \ldots, q-1\}$
\begin{align}
	\sum_{i = 0}^{q - 1} d_i(\theta') \int_0^\theta \left( \sum_{k=0}^{q - 1} \lambda_{i, k} \tau^k \right) \left( \sum_{k=0}^{q - 1} \rho_{j, k} \tau^k \right) \,\mathrm{d}\tau &= \sum_{k=0}^{q - 1} \lambda_{j, k} (\theta - \theta')^k \,.
	\label{eqn:delay_decoder_sys}
\end{align}
The integral can be evaluated in closed form. It holds
\begin{align*}
	\int_0^\theta \left( \sum_{k=0}^{q - 1} \lambda_{i, k} \tau^k \right) \left( \sum_{k=0}^{q - 1} \rho_{j, k} \tau^k \right) \,\mathrm{d}\tau
		&= \sum_{k=0}^{q - 1} \sum_{n=0}^{q - 1} \frac{\theta^{1 + n + k}}{1 + n + k} \rho_{j, n} \lambda_{i,k}
		&= (\vec \rho_j)^T \mat Q^q_\theta \vec \lambda_i\,,
\end{align*}
where $\vec \rho_j$ and $\vec \lambda_i$ are vectors of polynomial coefficients.
We can write \cref{eqn:delay_decoder_sys} in matrix-vector equation
\begin{align*}
	\mat P \mat Q^q_\theta \mat \Lambda^T \vec d (\theta') = \vec y^{\theta'}
\end{align*}
\marginnote{~~\\[-6em]This equation is implemented in the function \texttt{mk\_poly\_basis\_\\reencoder\_hilbert}.}%
where $\mat \Lambda, \mat P$ denote matrices of coefficients $\lambda_{i, k}$, $\rho_{j, k}$, respectively.
The vector $\vec y^{\theta'}$ is the right-hand side of \cref{eqn:delay_decoder_sys}.

Solving this system of equations tends to be numerically unstable.
This is partially due to the magnitude of the polynomial coefficients, and partially due to $\mat Q^q_\theta$.
For $\theta = 1$, this matrix is known as the \enquote{Hilbert matrix}, which is notoriously ill-conditioned---although a closed-form inverse exists (\cite[Section~2.8, p.~94]{press2007numerical}; \cite{choi1983tricks}).
For $q = 4$ the Hilbert matrix is
\begingroup
\renewcommand*{\arraystretch}{1.25}
\setlength\arraycolsep{8pt}
\begin{align*}
	\mat Q^4_1 &= \begin{pmatrix}
		\frac{1}{1} & \frac{1}{2} & \frac{1}{3} & \frac{1}{4}\\
		\frac{1}{2} & \frac{1}{3} & \frac{1}{4} & \frac{1}{5}\\
		\frac{1}{3} & \frac{1}{4} & \frac{1}{5} & \frac{1}{6}\\
		\frac{1}{4} & \frac{1}{5} & \frac{1}{6} & \frac{1}{7}
	\end{pmatrix} \,.
\end{align*}
\endgroup

\paragraph{Inverting the polynomial basis}
Above, we directly solved for a single delay decoder $\vec d(\theta')$.
Instead, we can compute an \enquote{inverse} of the $q$ polynomial basis functions.
Evaluating this inverse basis at any point $\theta'$ will result in the corresponding delay decoder.

The idea is simply that we would like to construct a set of polynomials $\tilde p_n$ with coefficients $\rho_{n, k}$ such that
\begin{align*}
	\langle p_i, \tilde p_j \rangle = \int_0^\theta \left( \sum_{k=0}^{q - 1} \lambda_{i, k} \tau^k \right) \left( \sum_{k=0}^{q - 1} \rho_{j, k} \tau^k \right) \,\mathrm{d}\tau = \delta_{ij} \,,
\end{align*}
where $\delta_{ij}$ is the Kronecker delta.
As above, the integral can be expressed using the scaled Hilbert matrix $\mat Q^q_\theta$.
We get the matrix-vector equation
\begin{align*}
	\mat P \mat Q^q_\theta \mat \Lambda^T = \mat \Lambda \mat Q^q_\theta \mat P^T = \mat I \,,
\end{align*}
\marginnote{~~\\[-7em]This equation is implemented in the functions 
\texttt{mk\_poly\_basis\\\_inverse} and \texttt{mk\_poly\_basis\_\\reencoder\_hilbert\_2}.}%
where, again, $\mat \Lambda$ and $\mat P$ denote matrices of polynomial coefficients $\lambda_{i, k}$, $\rho_{j, k}$, respectively.
This equation can be easily solved for the polynomial coefficients $\mat P$, which in turn define $\tilde p_n$.
Evaluating this set at any point $\theta'$ yields the corresponding delay decoder $\vec d(\theta)$.\footnote{We have not proved this formally, though showing this \emph{should} be relatively straight forward.}

Unfortunately, and unsurprisingly, this method tends to struggle with exactly the same numerical instabilities as the previous method.

\paragraph{Discrete inversion of the polynomial basis}
A rather naive approach that tends to work remarkably well is to perform the above basis inversion technique on a discretized basis.
At least under the assumption that the polynomials themselves can be evaluated at any point $t \in [0, \theta]$, this tends to be more stable.

Let $\mat L \in \mathbb{R}^{q \times N}$, where $N$ is the number of samples and
\begin{align*}
	(\mat L)_{ij} &= p_{i - 1}\left( \frac{\theta (j - 1)}{N - 1} \right) \,.
\end{align*}
The delay decoder corresponds to the individual rows of the pseudo-inverse $\mat L^+ \in \mathbb{R}^{N \times q}$ scaled by a factor $N / \theta$:
\begin{align*}
	\mat L^+ &= \mat L^T \big(\mat L \mat L^T\big)^{-1} \,, & \text{where } \vec d(\theta') &= \frac{N}{\theta} \mat L^+_{i} \; \text{ and } i = \left\lfloor \frac{(N - 1) \theta'}{\theta} \right\rfloor + 1 \,.
\end{align*}
This method provides reasonably precise results for $N$ on the order of $10^6$.

\begin{example}[The Chebyshev system delay decoder]
Using any of the above methods, we obtain the following delay decoder $\vec d(\theta)$ for the Chebyshev system and $q = 6$:
\begin{align*}
\vec d(\theta)   &= \begin{pmatrix}
	\cPos{4.79}&	\cPos{8.20}&	\cPos{9.23}&	\cPos{7.38}&	\cPos{8.12}&	\cPos{5.41}&	\cPos{5.87}\\
\end{pmatrix}^T \,.
\end{align*}
\marginnote{~~\\[-7em]This equation is implemented in the functions 
\texttt{mk\_poly\_basis\\\_rencoder} and \texttt{mk\_poly\_sys\\\_rencoder}.}%
The dampened Chebyshev system is depicted in \Cref{fig:cheb_damp_impulse}.
Note that in contrast to the Legendre polynomials, the individual delay decoder coefficients depend on the number of basis functions $q$.
This is generally the case for polynomial bases such as the Chebyshev basis that are not orthogonal with respect to a unit weighting.
While there seems to be some systematicity in the Chebyshev delay decoder with respect to $q$, it is not immediately apparent how to fit a precise closed-form solution to these results.

Note that polynomial bases that are orthogonal with respect to a unit weighting can be seen as a rotated version of the Legendre polynomials; there is little to be gained from not using the Legendre polynomials.
\end{example}

\section{Conclusion}

We presented an alternative derivation of the Legendre Delay Network (LDN) from an LTI system generating the Legendre polynomials.
We show that this system can be turned into the LDN by subtracting a delay re-encoder $\mat \Gamma$ from the feedback matrix $\mat A$.
This operation approximates a rectangle window applied to the impulse response.
This method can be used in conjunction with arbitrary polynomial bases, although numerical instabilities may make this difficult in practice.

Another downside of the presented method is that there is no formal guarantee that subtracting the delay re-encoder $\mat \Gamma$ from the feedback matrix $\mat A$ will actually result in an (almost) finite impulse response.
While this seems to be the case for the polynomial bases we tested---including random polynomial bases---it would be nice to have more formal guarantees.
Future work in this direction should establish a set of sufficient and necessary conditions pertaining the basis functions $f_n$ such that the dampened system $\mat A - \mat \Gamma$ generating these functions is guaranteed to decay to zero.

\section*{Acknowledgements}
We would like to thank Chris Eliasmith for his feedback on an earlier draft of this document and for supervising this work.
Furthermore, we would like to thank Aaron~R.~Voelker for sharing his encouraging thoughts on this new derivation.

\printbibliography

\section*{Revisions}
\begin{enumerate}[(1)]
 	\item February 9, 2021. First internal version.
 	\item February 26, 2021. Publication to arXiv.
 	\item March 9, 2021. Corrected some typos.
\end{enumerate}

\end{document}